\documentclass{svmult6}
\usepackage{graphicx}

\def\h{\ifmmode{h^{-1}{ \rm Mpc}}\else{$h^{-1}$ Mpc}\fi}
\renewcommand{\thefootnote}{\arabic{footnote}}

\begin{document}
\pagenumbering{arabic}
\chapter{Correlations at Large Scale}
\chapterauthors{M.J. Pons--Border{\'\i}a\footnote{Dept.
Matem{\'a}tica Aplicada y Estad{\'\i}stica, Univ. Polit{\'e}cnica de 
Cartagena (Spain)}, V.J. Mart{\'\i}nez\footnote{Observatori Astron{\`o}mic,
  Universitat de Val{\`e}ncia (Spain)}\\  B.
L{\'o}pez--Mart{\'\i} \footnote{Th{\"u}ringer Landessternwarte, 
Tautenburg (Germany)},  S. Paredes$^{1}$}
\shortauthname{Pons--Border{\'\i}a et al.}
\vspace*{-2.6cm}

\begin{abstract}  
  We show point processes generated 
  in different ways and having different
  structure, presenting very similar power-law two--point correlation
  functions at small scales and quite different shapes at large
  scales. 
\end{abstract}

The two-point correlation function $\xi(r)$ measures the excess 
probability ---with respect to a Poisson distribution--- of, given a 
point of a process, finding another point at a
distance $r$ of the first one ([2]). It is well-known that
$\xi(r)$, for the galaxy distribution, fits well a power
law at small scales ($r<10\h$). Here we analyze several point 
processes having similar power-law shapes at small scales, but different 
visual aspect. The differences are encapsulated
in the behavior of the correlation function at large scales as well as in
other statistical measures ([3]). The analyzed point processes are the 
following:

1. COX. A segment Cox process has been produced by randomly
scattering segments of length $l=10\h$ with a density $\lambda_s=0.0013$
within a cube of side 100\h, and then randomly
distributing points on the segments with density $\lambda_l=0.76923$ per unit
length. An analytical expression for $\xi(r)$ depending on these
parameters is known ([4]).

2. VORONOI. We have considered the vertices of a Voronoi tessellation ([5]) 
constructed from a binomial field with 1500 nuclei. There are 10085 vertices
(events of the point process) within a cube of sidelength
$100\sqrt{2}\h$. 
                                                    
3.  VIRGO. From a $\Lambda$-CDM N-body simulation of the Virgo Consortium,
a sample of simulated galaxies has been constructed by the GIF project ([1]). 
The sample contains $N = 15445$ galaxies within a cube of sidelength 
$141.3\h$.

\begin{figure}[t]
\includegraphics[width=\linewidth]{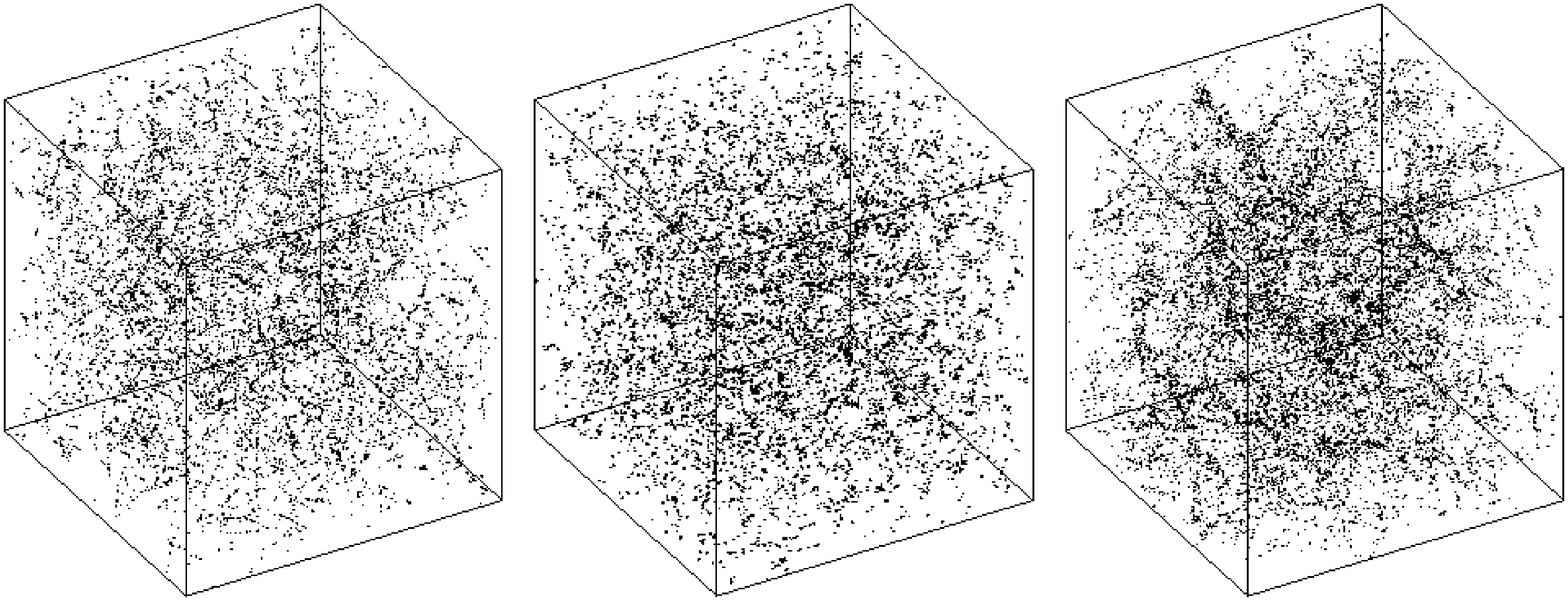}
\includegraphics[width=0.49\linewidth]{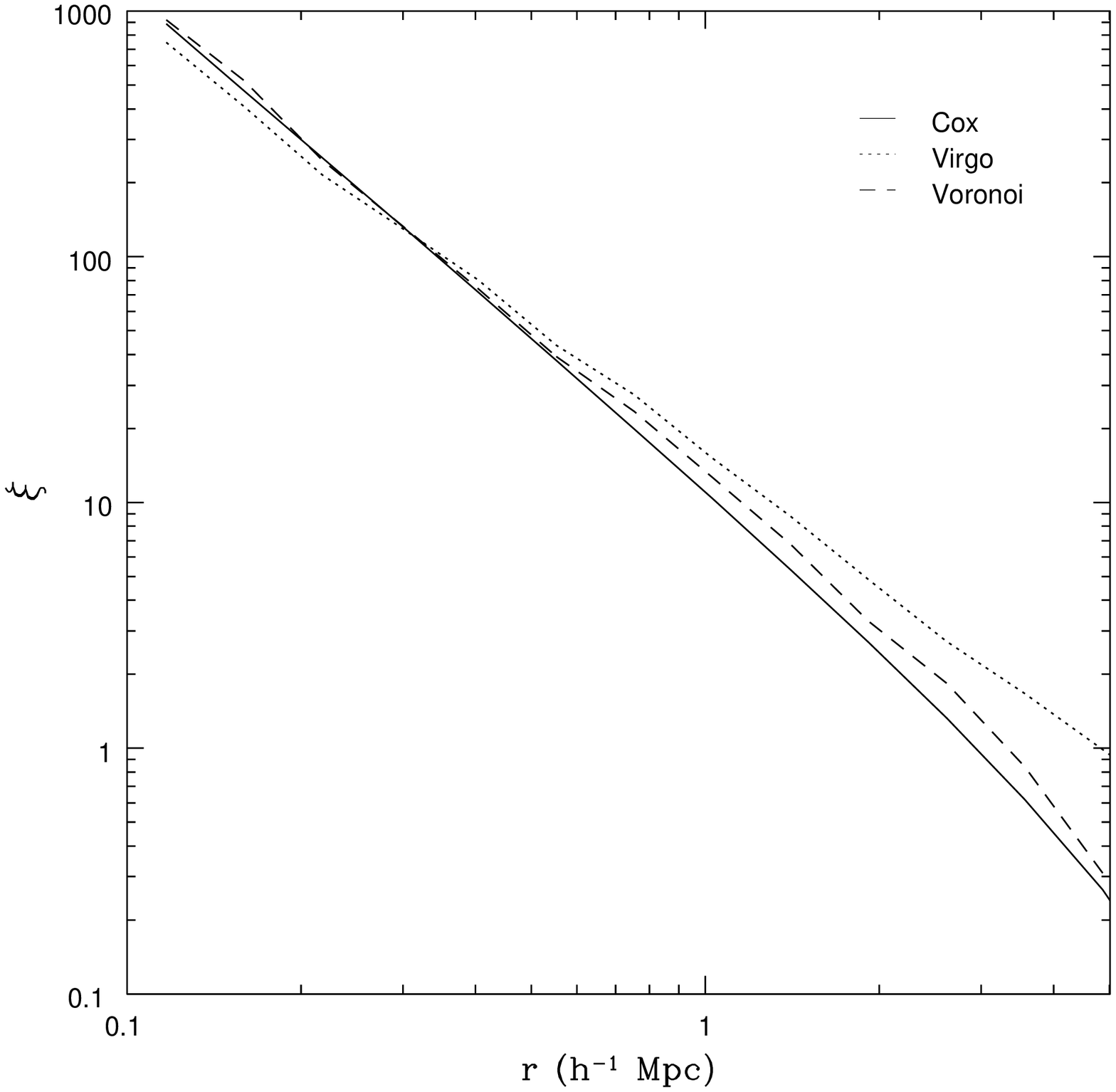}
\includegraphics[width=0.49\linewidth]{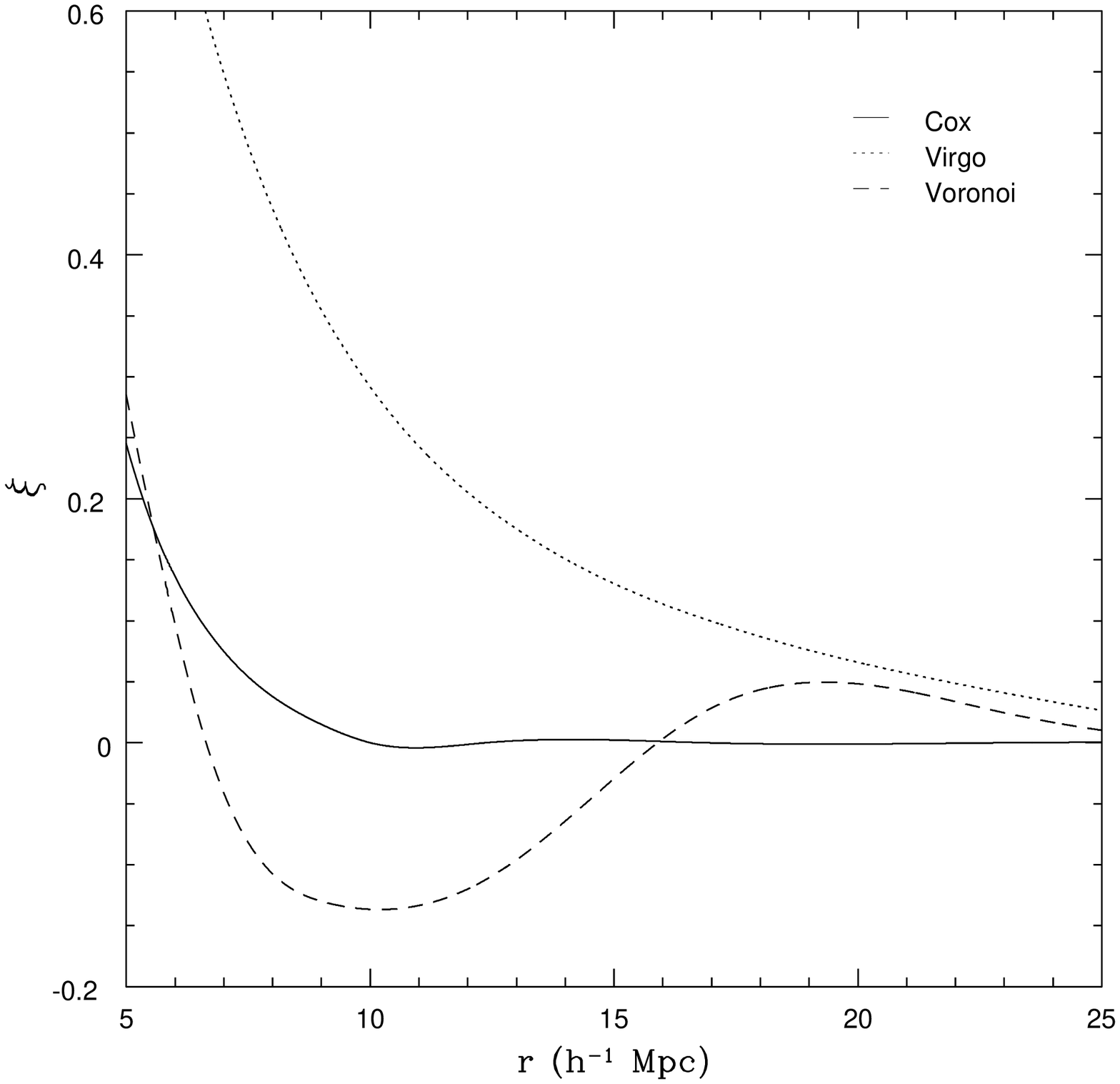}
\vspace*{-.1cm}\caption{Top: from left to right, the COX, VORONOI
and VIRGO point processes described in the text. Bottom: 
$\xi$ for the three processes at small scales (left panel) and at 
large scales (right panel).}
\end{figure}

At small scale the behavior of $\xi(r)$ is very
similar for the three clustering models --- power-law 
functions with comparable exponents. The
differences of the clustering properties of the three point processes
are better appreciated at large scale. For the Cox process $\xi(r)=0$
for $r\geq l$ whereas $\xi(r)$ for the N-body simulation $\xi(r)$ approaches zero more
gradually, taking place the first zero crossing at $\sim 30 \h$.
For the Voronoi vertices model, 
$\xi(r)$ behaves with damping oscillations around the zero value. 

We conclude by stressing that the behavior of $\xi$ at large scales
provide us with crucial information about the clustering properties
of point processes presenting similar power-law shapes at small scales.
Appropriate estimators had to be used to obtain this information, 
that can be complemented with other statistical measures ([3]).

\vspace*{-0.75cm}

\acknowledgments
This work was supported by the Spanish MCyT project AYA2000-2045.

\noindent
{\bf References}

\noindent
{[1]} Kauffmann, G. et al., 1999, MNRAS, 303, 188\\  
{[2]} Mart\'{\i}nez, V. J. and Saar, E., 2002, this volume. \\
{[3]} Mart\'{\i}nez, V. J. et al. 2002, in preparation. \\
{[4]} Pons-Border{\'\i}a, M.J. et al., 1999, ApJ, 523, 480\\ 
{[5]} van de Weygaert, R. and Icke, V., 1989, A\&A, 213, 1\\
\end{document}